\documentclass[11pt,twoside]{article}

\usepackage{asp2021}

\aspSuppressVolSlug
\resetcounters

\begin{document}

\title{The Universal Bayesian Imaging Kit}

\author{Torsten En{\ss}lin,$^{1-4}$ Vincent Eberle,$^{1,3}$ Matteo Guardiani,$^{1,3}$ and Margret Westerkamp$^{1,3}$}
\affil{$^1$Max Planck Institute for Astrophysics, Karl-Schwarzschild-Str. 1, 85748 Garching, Germany; \email{ensslin/veberle/matteani/margret@mpa-garching.mpg.de}}
\affil{$^2$Deutsches Zentrum für Astrophysik, Postplatz 1, 02826 Görlitz, Germany}
\affil{$^3$Ludwig-Maximilians-Universit\"at M\"unchen, Geschwister-Scholl-Platz 1, 80539 Munich, Germany}
\affil{$^4$Excellence Cluster ORIGINS, Boltzmannstr. 2, 85748 Garching, Germany}
\paperauthor{Sample~Author1}{Author1Email@email.edu}{ORCID_Or_Blank}{Author1 Institution}{Author1 Department}{City}{State/Province}{Postal Code}{Country}
\paperauthor{Sample~Author2}{Author2Email@email.edu}{ORCID_Or_Blank}{Author2 Institution}{Author2 Department}{City}{State/Province}{Postal Code}{Country}
\paperauthor{Sample~Author3}{Author3Email@email.edu}{ORCID_Or_Blank}{Author3 Institution}{Author3 Department}{City}{State/Province}{Postal Code}{Country}



\begin{abstract}
Bayesian imaging of astrophysical measurement data shares universal properties across the electromagnetic spectrum: it requires probabilistic descriptions of possible images and spectra, and instrument responses. To unify Bayesian imaging, we present the Universal Bayesian Imaging Kit (\texttt{UBIK}). 
Currently, \texttt{UBIK} images data from Chandra, eROSITA, JWST, and ALMA.
\texttt{UBIK} is based on information field theory (IFT), the mathematical theory of field inference, and on \texttt{NIFTy}, a package for numerical IFT.
\texttt{UBIK} provides sky models that are instrument independent and instrument interfaces that share common parts of their response representations. It is open source, can provide spatio-spectral image cubes, jointly analyses data from several instruments, and separates diffuse emission, point sources, and extended emission regions.
\end{abstract}



\section{Universality of imaging}

Currently, the different astronomical telescopes -- be these in the
radio, optical, X-ray, or any other wavelength regime -- come with
their specific imaging software. The task these systems address, however,
is always the same, to convert the instrument data into an image of
the sky while removing imperfection of the measurement process as
far as possible. The data the telescopes deliver is never sufficient
to specify the sky fully, there are always limitations due to various instrument effects.
Thus, any generated image is a reconstruction of
the celestial reality, partly uncertain itself and based on assumptions
on the sky and the measurement process. A proper accounting of uncertainties
is only possible in a strict probabilistic setting. 
This requires that the instrument and its statistical noise properties as well as the exploited prior knowledge are explicitly specified and used. 
Thus, imaging should be Bayesian. 
Since from an epistemological perspective no principal
difference exists between the different imaging problems -- in all
data is converted into an image of the world using auxiliary knowledge
-- imaging should also be universal, meaning that the same methods
are used across all measurement modalities.

\section{Universal Bayesian Imaging Kit}

\begin{figure}[t]
	\includegraphics[width=1\columnwidth]{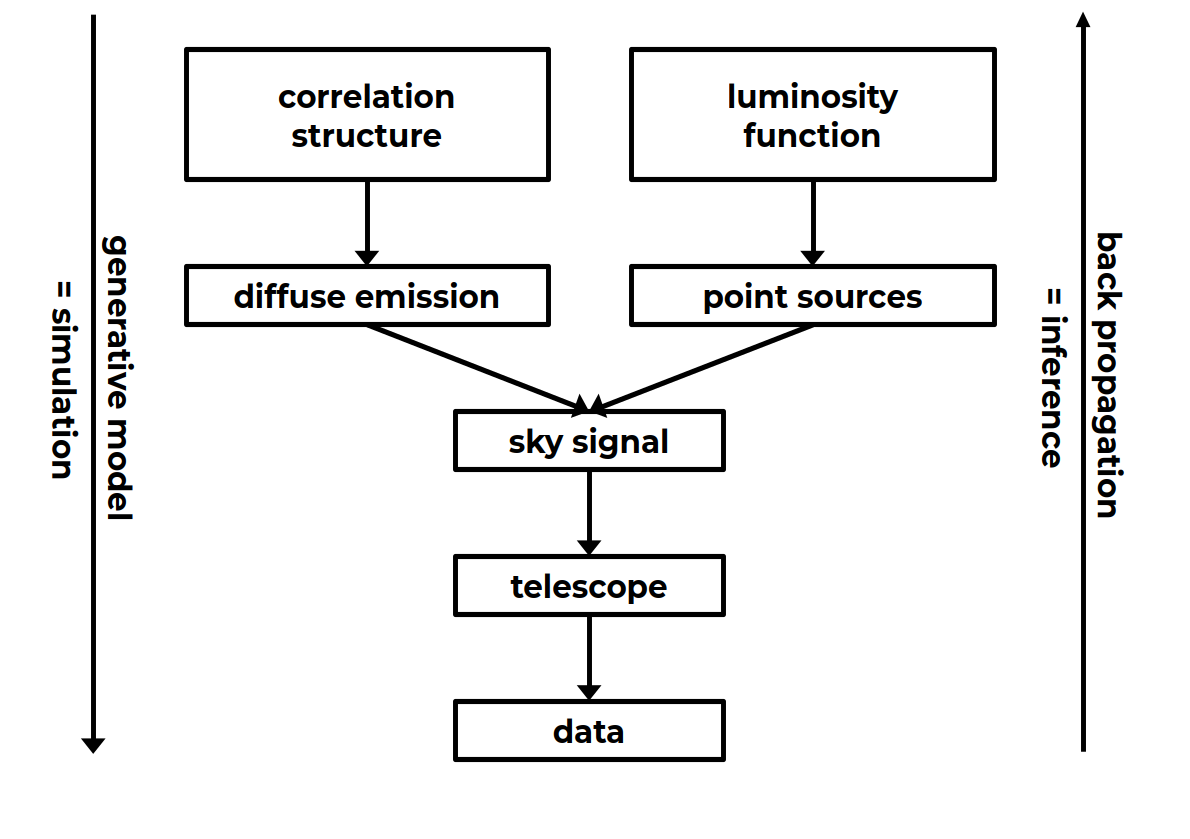}\caption{Graphical representation of a typical generative model used in astronomical
		imaging with \texttt{UBIK}.\protect\label{fig:generative-model}}
\end{figure}
For the aim of a unified approach to Bayesian imaging, we present \texttt{UBIK}, the \textbf{U}niversal \textbf{B}ayesian
\textbf{I}maging \textbf{K}it \citep{2024arXiv240910381E}, a Python-based software system (\url{https://github.com/NIFTy-PPL/J-UBIK}) that
unifies imaging for a growing number of instruments and aims at enabling multi-instrument imaging and cross-calibration. 

\texttt{UBIK}
permits the user to describe the measurement situation in terms of
a generative forward model, see Fig.\ \ref{fig:generative-model}.
The model can be built out of a number of preexisting building blocks
and user supplied elements. For example, the model shown in Fig.\ \ref{fig:generative-model}
might be used to image data from an X-ray telescope like Chandra.
For this it consists of the following elements:
\begin{itemize}
	\item A latent layer (not shown), which is the start of the forward model
	and is assumed to contain a priori independent standard Gaussian random
	variables, collectively described by a vector $\xi\in\mathbb{R}^{n},$ with $\mathcal{P}(\xi)=\mathcal{G}(\xi,1\!\!1)$ that sources all variations of the model realizations.
	\item Transformations that shape these latent variables into model components
	$s=f(\xi)$ with the desired a priori statistics $\mathcal{P}(s)=\mathcal{P}(\xi)\,||\frac{\partial f(\xi)}{\partial\xi}||_{\xi=f^{-1}(s)}^{-1}$.
	These comprise the unknown spatial or spectral correlation function
	of the diffuse sky emission field, properties of the luminosity function
	of point sources, the diffuse and point source fields, calibration
	properties, and so on.
	\item A likelihood function $\mathcal{P}(d|s)$ that can be used to generate
	synthetic data $d$ and that allows a comparison of expected detector
	imprint of an assumed sky signal with the observed data in light of
	the instrument noise statistics. This function needs to describe all
	relevant effects that imprint on the data, including eventually spatially
	varying point spread functions, in-homogeneous sky exposure, Poisson,
	Gaussian, or other noise processes, calibration fluctuations, cosmic
	ray hits, and more.
\end{itemize}
\texttt{UBIK} can run the model in forward direction and thereby generate
synthetic skies and corresponding data sets. These can be used to
demonstrate, verify, and characterize the imaging. The imaging itself
is performed with real or simulated data entered at the end of the
model. This data is compared to the expectation value of data for
the model given the current state of the latent (and other) variable
on the basis of the latter's log-likelihood. The discrepancy between
real and expected data is back propagated through the generative process
into the initial latent variable layer of the model. As this inversion
is typically not unique -- there exist a manifold of possible latent
space variable configurations that are consistent with the data within
the noise statistics -- it is performed in a probabilistic fashion.
This means that a sample of possible images are generated, drawn approximately
from the Bayesian posterior distribution $\mathcal{P}(s|d)=\mathcal{P}(\xi|d)\,||\frac{\partial f(\xi)}{\partial\xi}||_{\xi=f^{-1}(s)}^{-1}$.

\section{Variational inference}

Since imaging involves the determination of many unknowns, the field
values at all pixel locations as well as many parameters of the higher
level of the generative model, the space of potential images is very
high dimensional. In order to represent the posterior distribution
over this space, \texttt{UBIK} usually resorts to variational inference
(VI) as implemented in the \texttt{NIFTy} package.\footnote{\texttt{NIFTy} stands for \emph{Numerical Information Field Theory}
\citep{2013A&A...554A..26S,2019ascl.soft03008A,2024JOSS....9.6593E}. It uses \texttt{JAX} for just in time compilation, GPU access, and automatic differentiation. \emph{Information Field Theory} \citep[IFT,][]{2009PhRvD..80j5005E} is information theory for fields, the mathematical framework of Bayesian inference for field-like quantities.} 
There are two VI schemes available, \emph{Metric Gaussian VI}
\citep[MGVI,][]{2019arXiv190111033K} and \emph{geometric VI} \citep[geoVI,][]{2021Entrp..23..853F}. 

The former scheme, MGVI, approximates the posterior distribution by
a Gaussian distribution $\mathcal{P}(\xi|\overline{\xi})=\mathcal{G}(\xi-\overline{\xi},\Xi(\overline{\xi}))$.
The mean of the Gaussian, $\overline{\xi}$, contains the variational
parameters that are optimized by minimizing the variational cross
entropy
$
\mathcal{D}(\overline{\xi})=\int\text{d}\xi\,\mathcal{P}(\xi|\overline{\xi})\,\ln\frac{\mathcal{P}(\xi|\overline{\xi})}{\mathcal{P}(\xi|d)}.
$
The covariance $\Xi(\overline{\xi})$ of this Gaussian is approximated
with the help of the Fisher information matrix
$
	M_{\overline{\xi}}=\left\langle \frac{\partial\ln\mathcal{P}(\xi|d)}{\partial\xi}\,\frac{\partial\ln\mathcal{P}(\xi|d)}{\partial\xi}^{\dagger}\right\rangle _{(d|\overline{\xi})}
$
as
$
	\Xi(\overline{\xi})=\left[1\!\!1+M_{\overline{\xi}}\right]^{-1}.
$
This form ensures that for directions in the latent space for which
the posterior is Gaussian or not affected by the data, the covariance
is exact. If this is true for all directions, the Gaussian posterior
approximation is exact as well. 

The latter scheme, geoVI, improves on this by constructing a further
latent space that is reachable via an approximate normalizing transformation,
$y=g_{\overline{\xi}}(\xi)$. This is done by constructing approximate Riemann normal
coordinates according to the metric $1\!\!1+M_{\xi}$ starting from
an expansion point $\overline{\xi}$. The optimal expansion point
is found by minimizing $\mathcal{D}(\overline{\xi})$ 
for the approximate distribution $\mathcal{P}(\xi|\overline{\xi})=\mathcal{G}(g_{\overline{\xi}}(\xi),1\!\!1)\,||\frac{\partial g_{\overline{\xi}}(\xi)}{\partial\xi}||^{-1}$.

In both schemes, MGVI and geoVI, initially Gaussian samples are converted
non-linearly into the model parameters, either via $s=f(\xi)$ in
MGVI or with $s=f(g_{\overline{\xi}}^{-1}(y))$ in geoVI. These samples can then be
used to calculate any kind of posterior summary statistics via
$	\left\langle F(s)\right\rangle _{(s|d)}\approx\frac{1}{N}\sum_{i=1}^{N}F(s_{i}),
$
where $F(s)$ denotes the desired quantity and $s_{i}$ any of the
$N$ samples.

\section{\texttt{UBIK} components}

\texttt{UBIK} provides
model components to represent instrument response functions and their noise statistics and to describe various signals over one or higher dimensional domains like space, time, energy, and combinations thereof. 
It comes with ready to use instrument descriptions for a growing number of instruments, currently Chandra, eROSITA, JWST, and ALMA \citep{2024A&A...682A.146R,2024A&A...684A.155W,2024arXiv241014599E,2025A&A...703A.203G}.


\acknowledgements M.G., V.E., and M.W. acknowledge financial support from the German Aerospace Center and Federal Ministry of Education and Research through the project Universal Bayesian Imaging Kit -- Information Field Theory for Space Instrumentation (Förderkennzeichen 50OO2103). M.G. acknowledges support from the European Union-funded project mw-atlas under grant agreement No.~101166905. V.E. acknowledges funding through the German Federal Ministry of Education and Research for the project ErUM-IFT: Informationsfeldtheorie für Experimente an Großforschungsanlagen (Förderkennzeichen: 05D23EO1).

\bibliography{374}  


\end{document}